\documentstyle[emulateapj, psfig]{article}

\lefthead{COORAY \& HUTERER}
\righthead{GRAVITATIONAL LENSING AS A PROBE OF QUINTESSENCE}

\begin{document}
\title{Gravitational Lensing as a Probe of Quintessence}
\author{Asantha R. Cooray$^1$, Dragan Huterer$^2$}
\affil{$^1$Department of Astronomy and Astrophysics,
University of Chicago, Chicago IL 60637\\
$^2$Department of Physics, University of Chicago, Chicago IL 60637\\
E-mail: asante@hyde.uchicago.edu, dhuterer@aether.uchicago.edu}

\begin{center}
{\it Accepted for publication in The Astrophysical Journal Letters}
\end{center}

\received{\today}
\revised{}
\accepted{}


\begin{abstract}

A large number of cosmological studies now suggest that roughly two-thirds
of the critical energy density of the Universe exists in a component with
negative pressure. If the equation of state of such an energy component
varies with time, it should in principle be possible to identify such
a variation using cosmological probes over a wide range in redshift.  Proper
detection of any time variation, however, requires cosmological probes
beyond the currently studied range in redshift of $\sim$ 0.1 to 1.  We
extend our analysis to gravitational lensing statistics at high redshift
and suggest that a reliable sample of lensed sources, out to a redshift of
$\sim$ 5, can be used to constrain the variation of the equation of state,
provided that both the redshift distribution of lensed sources and the
selection function involved with the lensed source discovery process are
known.  An exciting opportunity to catalog an adequate sample of lensed
sources (quasars) to probe quintessence is now available
with the ongoing Sloan Digital Sky Survey. Writing $w(z)\approx
w_{\scriptscriptstyle 0} + z (dw/dz)_{\scriptscriptstyle 0}$, we study the
expected accuracy to which the equation of state today
$w_{\scriptscriptstyle 0}$ and its rate of change
$(dw/dz)_{\scriptscriptstyle 0}$ can simultaneously be constrained.  Such a
determination can rule out some missing-energy candidates, such as classes
of quintessence models or a cosmological constant.

\end{abstract}



\section{Introduction}

There is now strong evidence that the matter density of the Universe, both
baryonic and dark, is smaller than the critical value predicted by
inflation. Some of the same cosmological probes that suggest the ratio of
matter density to critical density is less than one prefer the presence of
an additional component with energy density such that the total energy
density of the Universe is in fact the critical value (e.g., Lineweaver
1998; Perlmutter et al. 1998; Riess et al. 1998). Type Ia
supernovae and other observations have recently provided evidence that the
equation of state, $w$, of this component is $-1 \lesssim w
\equiv P/\rho \lesssim -1/3$ (Garnavich et al.  1998; Waga \& Miceli 1998;
Cooray 1999).

The oldest known candidate for this additional energy is the cosmological
constant, characterized by $w=-1$.  However, other possibilities have now
been considered, including a slowly-varying, spatially inhomogeneous scalar
field component better known as quintessence (e.g., Huey et al. 1998;
Zlatev et al. 1998; Wang \& Steinhardt 1998).  Certain quintessence models,
called {\it tracker models} (Ferreira \& Joyce 1997; Zlatev et al. 1998;
see Steinhardt et al. 1998 for a comprehensive review),
have the feature that the energy density of the scalar field, $\Omega_Q$,
is a fixed fraction of the energy density of the dominant
component. Therefore such models may explain the coincidence problem ---
why $\Omega_Q$ is of the same order of magnitude as $\Omega_M$ today.  In
these tracker models $w$ is a function of redshift, and typically varies
from $w\sim 1/3$ during the radiation dominated era to $w\sim -0.2$ during
the matter-dominated era and finally to a value $w\sim -0.8$ during late
epochs (today).  Time variation of the equation of state is even more
prominent in scalar field models involving pseudo-Nambu-Goldstone bosons
(PNGB models) (Frieman et al. 1995).
In such models the field is frozen to its initial value due
to the large expansion rate, but becomes dynamical at some later stage
at redshift $z_x$.  Likely values for $z_x$ are roughly between 3 and 0
(Coble et al. 1997), which means that interesting dynamics --- and hence the
variation in the equation of state --- happen at redshift of a few.  Huterer
\& Turner (1999) point out that distance measurements to Type Ia supernovae
offer a possibility to reconstruct the quintessence potential, while
Starobinsky (1998) and Wang \& Steinhardt (1998)
suggested the possibility of using cluster abundances as
a function of redshift as well.

In \S~2, we study constraints on $w(z)$ based on current Type Ia supernovae
distances, gravitational lensing statistics, and globular cluster ages.  In
\S~3, we consider the possibility of imposing
reliable constraints on $w(z)$ based
on cosmological probes at high redshift, in particular gravitational
lensing statistics.  We follow the conventions that the Hubble constant,
$H_0$, is 100\,$h$\ km~s$^{-1}$~Mpc$^{-1}$, the present matter energy
density in units of the closure density is $\Omega_M$, and the normalized
present day energy density in the unknown component is $\Omega_Q$.

\section{Current Constraints on $w(z)$}

Since a generic quintessence model has a time-varying equation of state,
one should be able to distinguish it from models where the equation of
state is time independent (e.g, Turner \& White 1997).  In this {\it
Letter}, we write the equation of state of the unknown component as
\begin{equation}
w(z)\approx w_{\scriptscriptstyle 0}+z(dw/dz)_{\scriptscriptstyle 0}.
\end{equation}
This relation should be a good approximation for most quintessence models
out to redshift of a few and, of course, exact for models where $w(z)$ is a
constant or changing slowly. Note that negative
$(dw/dz)_{\scriptscriptstyle 0}$ corresponds to an equation of state which
is larger today compared to early epochs. Models where the scalar field
is initially frozen typically exhibit such behavior, while for tracker
field models $(dw/dz)_{\scriptscriptstyle 0} >0$.

In order to constrain $w(z)$, we extend current published analyses which
have so far considered the existence of a redshift-independent equation of
state (e.g., Cooray 1999; Garnavich et al. 1998; Waga \& Miceli
1998). Since the only difference between this study and previous ones is
that we now allow $w$ to vary with $z$, formalisms presented in previous
papers should also hold except for the fact that $w$ is now redshift
dependent.  We refer the reader to previous work for detailed formulae and
calculational methods.

\vskip 2mm
\hbox{~}

\centerline{\psfig{file=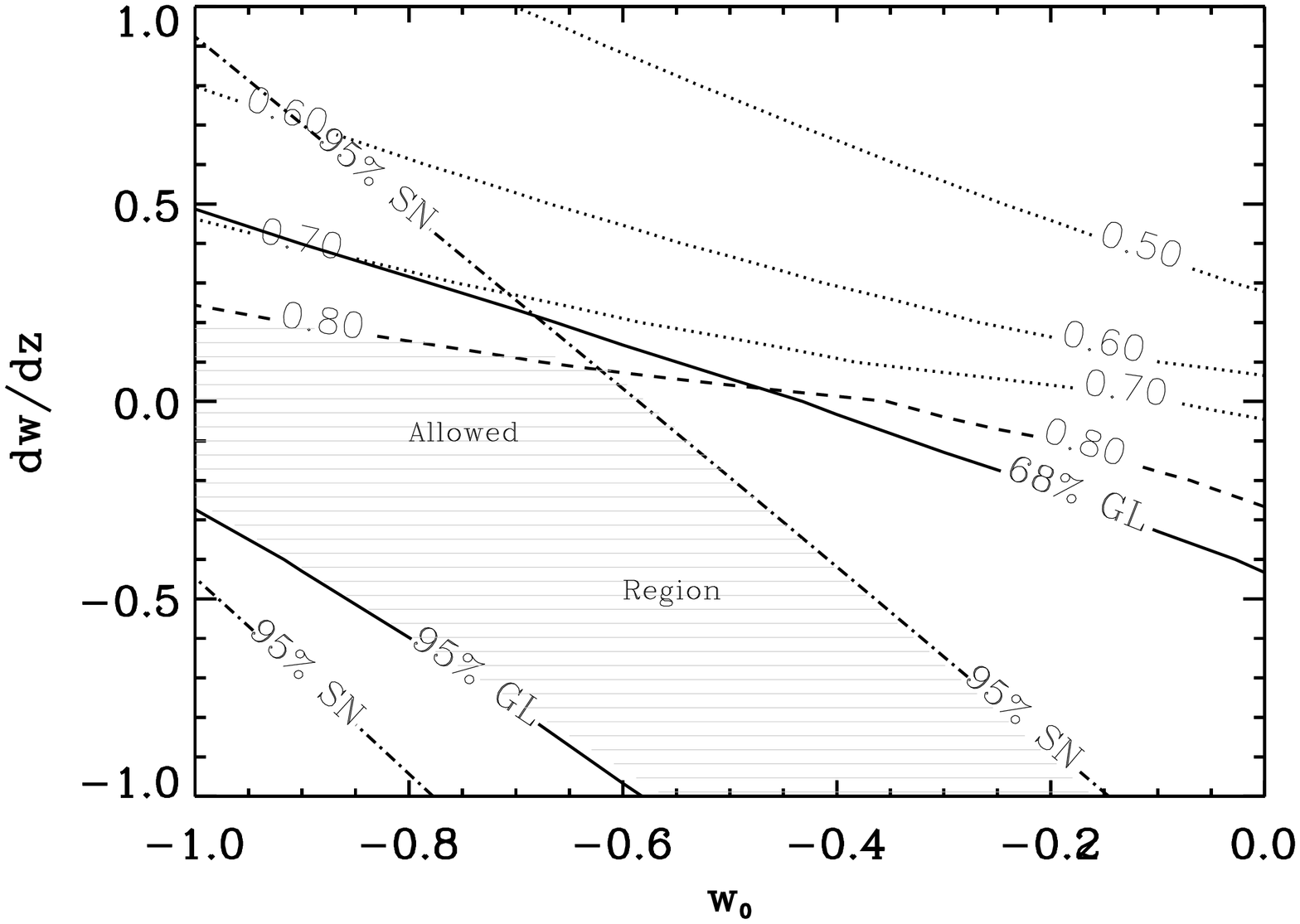,width=3.7in,angle=90}}
\noindent{
\scriptsize \addtolength{\baselineskip}{-3pt}
\vskip 1mm

\begin{normalsize}
Fig.~1.\ Current constraints on $w(z) \approx w_{\scriptscriptstyle 0}
+ z (dw/dz)_{\scriptscriptstyle  0}$.
The {\it dot-dashed} lines are the current type Ia supernovae
(Riess et al. 1998; Perlmutter et al. 1998)
constraints at the 95\% confidence level,
while the upper limits from gravitational lensing statistics
(Cooray 1999) is shown with {\it solid} lines. The age of the Universe
as a function of $H_0t_0$ is show by {\it dotted} lines.
A conservative lower limit on this parameter is 0.8
({\it dashed} line) when age of the
Universe is $\sim$ 14 to 15 Gyr and $H_0 \sim 65$ km
s$^{-1}$ Mpc$^{-1}$.  The {\it shaded} region defines the
parameter region allowed by current data.
\end{normalsize}

\vskip 3mm
\addtolength{\baselineskip}{3pt}
}

In Fig.~1 we summarize current constraints on $w(z)$ as given in
Equation~1.  Here we have assumed a flat Universe with $\Omega_M=1/3$.
Evidence for such low matter density, independent of the nature of the
additional energy component, comes primarily from the mass density within
galaxy clusters (e.g., Evrard 1997).  As shown in Fig~1, there is a wide
range of possibilities for $w(z)$, and the allowed parameter space is
consistent with the degeneracies discussed in literature (e.g., Zlatev et
al. 1998).  Even though the $w \sim -1/3$ model has been ruled out by
combined type Ia supernovae, gravitational lensing and globular cluster
ages, we now note that a model in which $w_{\scriptscriptstyle 0} \sim
-1/3$ but $(dw/dz)_{\scriptscriptstyle 0}
\sim -0.9$ is fully consistent with the current observational data.

In order to test whether one can constrain $w(z)$ better than the current
data, we increased the Type Ia supernova sample between redshifts of 0.1 to
1; however, the degeneracy between $w_{\scriptscriptstyle 0}$ and
$(dw/dz)_{\scriptscriptstyle 0}$ did not change appreciably.  Increasing
the upper redshift of supernova samples decreased the degeneracies; thus,
cosmological probes at high redshifts are needed to properly distinguish
redshift-dependent $w(z)$ from a constant $w$ model.  A probe to a much
higher redshift is provided by the CMB anisotropy data; however, as pointed
out in Huterer
\& Turner (1998) and Huey et al. (1998), CMB anisotropy is not a strong
probe of $w(z)$. This is due to the fact that $w(z)$ affects mostly the
lower multipoles, which cannot be measured precisely due to cosmic
variance.  Also, supernovae and galaxy clusters are
unlikely to constrain $w(z)$ in the near
future given that current observational programs are not likely to
recover them at high redshifts ($z \gtrsim 2$).

\vskip 2mm
\hbox{~}

\centerline{\psfig{file=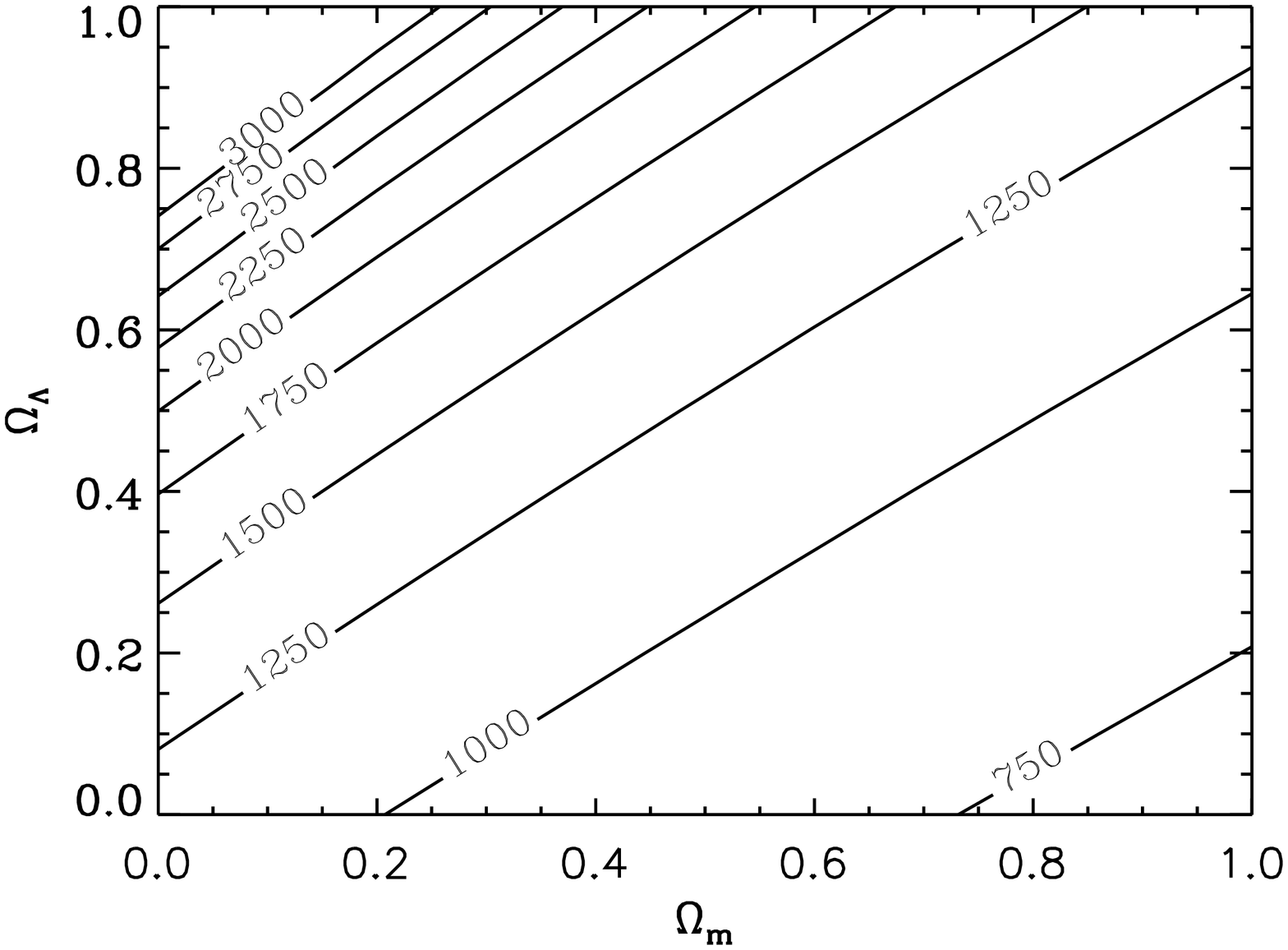,width=3.7in,angle=90}}
\noindent{
\scriptsize \addtolength{\baselineskip}{-3pt}
\vskip 1mm

\begin{normalsize}
Fig.~2.\ Expected number of multiply imaged quasars with
image separations between 1 and 6 arcsecs in the SDSS data.
We have only counted lensed quasars with at least two images greater than
a magnitude limit of 21.
The expected number is a constant along the
$\Omega_M-\Omega_\Lambda$ lines. For $\Omega_M=0.3$ and
$\Omega_\Lambda=0.7$, we expect $\sim$ 2000 lensed
sources to be detected within the SDSS.
\end{normalsize}

\vskip 3mm
\addtolength{\baselineskip}{3pt}
}
Thus, an alternative probe to high redshifts is needed. In the next section
we consider the possibility of using gravitational lensing statistics,
in particular the redshift distribution of strongly lensed sources, as a probe
of $w(z)$. Such statistics in principle probe the volume of the
Universe out to a redshift of $\sim$ 5.  In the past, lensing statistics
were hampered by the lack of large samples of lensed sources with a well
known selection function and their redshift distributions, which are all
needed to constrain $w(z)$. An exciting possibility is now provided by the
upcoming high-quality data from the Sloan Digital Sky
Survey\footnote{http://www.sdss.org} (SDSS; Gunn \& Knapp 1993) which
is going to image $\pi$ steradians of the sky down to a 1-$\sigma$
magnitude limit of $\sim$ 23.
Since no high redshift ($z \gtrsim 2$) cosmological probes yet
exist, gravitational lensing statistics may be the prime
 candidate to study $w(z)$.

\section{Gravitational Lensing Statistics}

In order to calculate the expected number of lensed quasars, in particular
considering the SDSS, we extend previous calculations in Wallington \&
Narayan (1993; also, Dodelson \& MacMinn 1997) and Cheng \& Krauss
(1998). Our calculation follows that of Cooray et al. (1999) in which we
calculated the number of lensed galaxies in the Hubble Deep Field.  We
follow the magnification bias (e.g., Kochanek 1991) calculation in Cheng \&
Krauss (1998). We calculate the expected number of lensed sources as a
function of the magnitude limit of the SDSS and select sources with image
separations between 1 and 6 arcsecs. This range is selected based on image
resolution and source confusion limits.  Our prediction, shown in Fig.~2 as
a function of $\Omega_M$ and $\Omega_\Lambda$, is based on current
determination of the quasar luminosity function, which is likely to be
updated once adequate quasars statistics are available from the SDSS.

As shown, there are about 2000 lensed quasars down to a limiting magnitude
of 21 that could in principle be detected from the SDSS data if
$\Omega_M=0.3$ and $\Omega_\Lambda=0.7$. This number drops to about $\sim$
600 when $\Omega_M=1.0$; however, such a cosmological model is already
ruled out by current data. In this calculation, we have ignored effects due
to extinction and dust; this is an issue with no consensus among various
studies (Malhotra et al. 1997; Falco et al. 1998). It is likely that a
combined analysis of statistics from ongoing lensed radio sources and
optical searches may increase knowledge on such systematic effects.  The
SDSS lensed quasars lie in redshifts out to $z \sim$ 5. Detection of such
high redshift lensed sources is likely to be aided by the 5 color imaging data,
including the $z$-filter. This possibility has already been demonstrated by
the detection of some of the highest redshift quasars known today using the
SDSS first year test images (Fan et al. 1998).

\section{Lensing Constraints on $w(z)$: Prospects}

In order to estimate the accuracy to which one can constrain $w(z)$ based
on gravitational lensing statistics, we take a Fisher matrix approach with
parameters $w_{\scriptscriptstyle 0}$ and $(dw/dz)_{\scriptscriptstyle 0}$,
and assume a flat universe.  As stated in literature (e.g., Tegmark et
al. 1997), Fisher matrix analysis allows one to estimate the best
statistical errors on parameters calculated from a given data set. The
Fisher matrix $F$ is given by:

\begin{equation}
F_{ij} = -\left< \partial^2 \ln L \over \partial p_i \partial p_j
 \right>_{\bf x},
\end{equation}
where $L$ is the likelihood of observing data set ${\bf x}$
given the parameters $p_1 \ldots p_n$.

We bin the observations (number of lensed quasars) in redshift bins of
width $\Delta z$ out to a maximum redshift $z_{\rm max}$. Since the
selection function for quasar discovery process is still unknown, we adopt
a Poisson likelihood function at each redshift bin $\Delta z$ centered at
redshift $z$.  The expected number of lensed sources, $N_{\rm exp}$, in
each redshift bin takes into account the magnitude limit, range of allowed
image separations as well as the magnification bias.  With these
approximations, we can now write the Fisher matrix as:

\begin{equation}
F_{ij} = \sum_{\Delta z}\frac{1}{N_{\rm exp}(z, \Delta z)}
               {\partial N_{\rm exp}(z, \Delta z) \over \partial p_i}
        {\partial N_{\rm exp}(z, \Delta z) \over \partial p_j}\,.
\label{eqn:Fisher}
\end{equation}
This form for the Fisher matrix is identical to the one derived from
a Gaussian distribution when the uncertainty ($\sigma$) of the distribution
is taken to be equal to the shot-noise term $\sqrt{N_{\rm exp}(z, \Delta z)}$.

\vskip 2mm
\hbox{~}

\centerline{\psfig{file=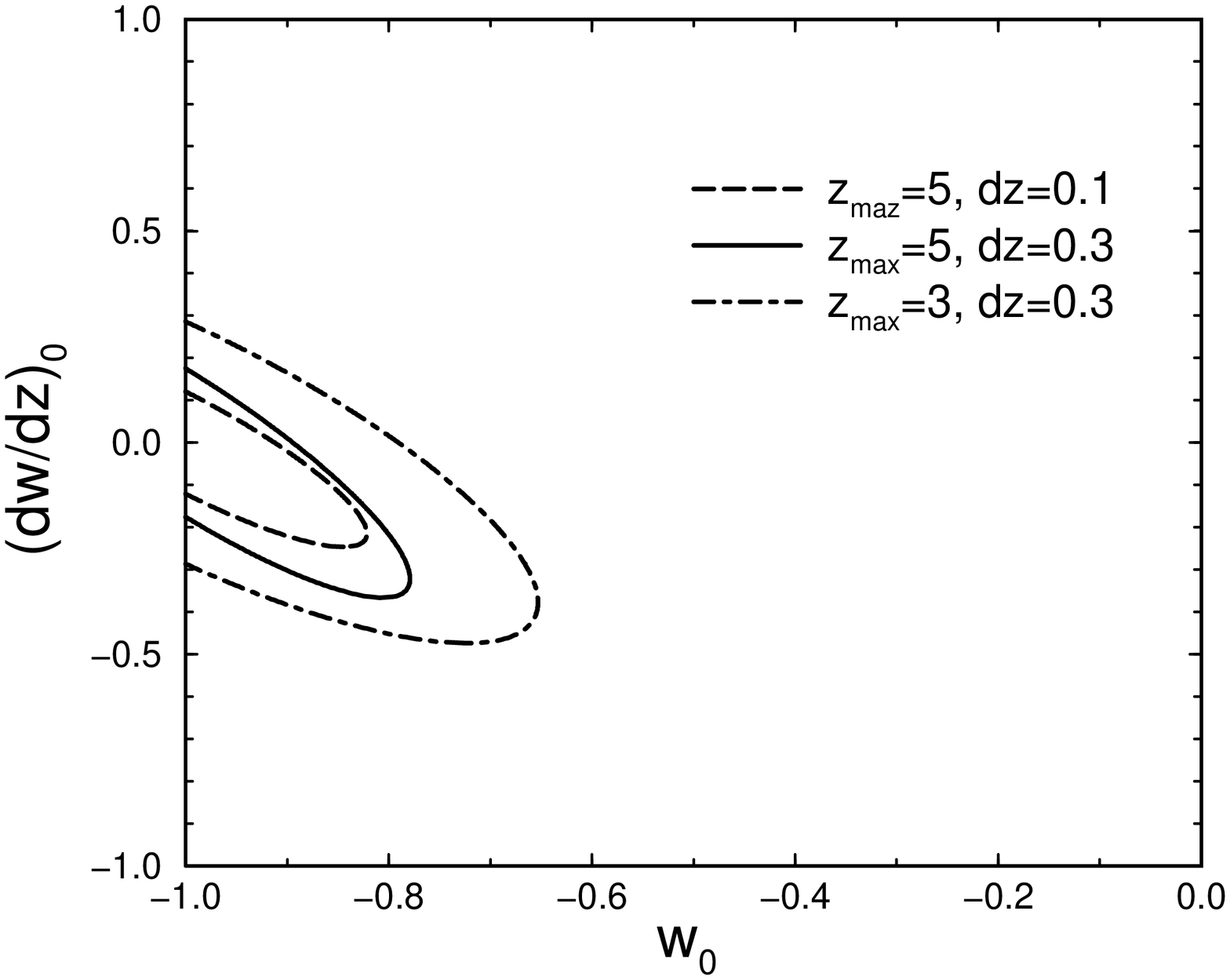,width=3.7in,angle=-90}}
\noindent{
\scriptsize \addtolength{\baselineskip}{-3pt}
\vskip 1mm

\begin{normalsize}
Fig.~3.\ Constraints  (2-$\sigma$) on $w_{\scriptscriptstyle 0}$ and
$(dw/dz)_{\scriptscriptstyle 0}$ using lensed source redshift distribution
expected from the SDSS for a fiducial cosmological model of
$w_{\scriptscriptstyle 0}=-1$, $(dw/dz)_{\scriptscriptstyle 0}=0$, and
marginalised over $\Omega_M$ ($\Omega_M=0.3 \pm 0.1$ and
$\Omega_Q=1-\Omega_M$). We show the expected 2$\sigma$ errors when
the redshift distribution of lensed sources is known with an accuracy of
0.1 and 0.3 while the maximum redshift probed by lensing statistics is 3 and
5.
\end{normalsize}
\vskip 3mm
\addtolength{\baselineskip}{3pt}
}

In Fig.~3 we show the expected (2-$\sigma$) uncertainties in
$w_{\scriptscriptstyle 0}$ and $(dw/dz)_{\scriptscriptstyle 0}$.  Here we
considered a flat universe with three parameters, $\Omega_M$,
$w_{\scriptscriptstyle 0}$, and $(dw/dz)_{\scriptscriptstyle 0}$. We
marginalised over $\Omega_M$ allowing $\Omega_M= 0.3 \pm 0.1$ (1-$\sigma$)
and considered a fiducial model with $w_{\scriptscriptstyle 0}=-1$ and
$(dw/dz)_{\scriptscriptstyle 0}=0$. The three curves in Fig.~3 show
variation of the constraint region with the redshift bin width $\Delta z$
(which is roughly equal to the uncertainty in the redshift determination)
and with the maximum redshift of detected quasars $z_{\rm max}$.
Photometric redshifts now allow redshift determinations with an accuracy of
order 0.1 68\% of the time and of order 0.3 100\% of the time (e.g.,
Hogg et al. 1998).  It is apparent that the size of the constraint region
decreases significantly with increasing $z_{\rm max}$ and decreasing
$\Delta z$.  Note that one can quite safely assume that $z_{\rm max}\approx
5$ for quasars in a survey such as the SDSS.  Additionally, these
calculations assume a limiting magnitude of 21, much lower than the
expected 1-$ \sigma$ limiting magnitude of 23 for the SDSS data.

\vskip 2mm
\hbox{~}

\centerline{\psfig{file=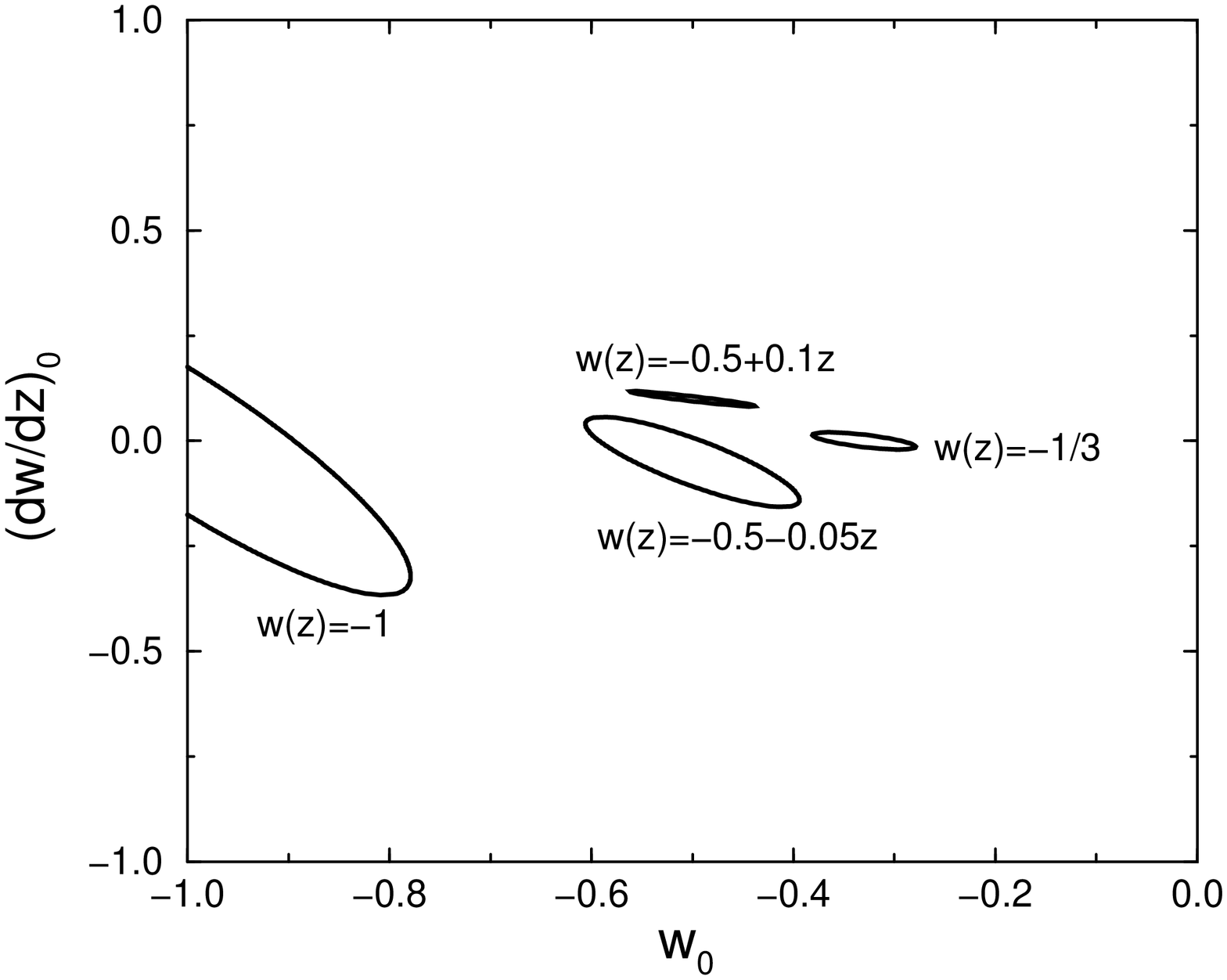,width=3.7in,angle=-90}}
\noindent{
\scriptsize \addtolength{\baselineskip}{-3pt}
\vskip 1mm
\begin{normalsize}
Fig.~4.\ Constraints  (2-$\sigma$) on $w_{\scriptscriptstyle 0}$ and
$(dw/dz)_{\scriptscriptstyle 0}$ using lensed source redshift distribution
expected from the SDSS for four fiducial cosmological models when redshift
of lensed sources is known with an accuracy of 0.3 out to a redshift of 5.
In all cases we assumed a flat universe and marginalised over $\Omega_M$
($\Omega_M=0.3 \pm 0.1$).
\end{normalsize}
\vskip 3mm
\addtolength{\baselineskip}{3pt}
}

Fig.~4 shows constraints in the $w_{\scriptscriptstyle 0}$ --
$(dw/dz)_{\scriptscriptstyle 0}$ plane for four fiducial models. Models
shown are the cosmological constant ($w(z)=-1$), non-Abelian cosmic strings
($w(z)=-1/3$; Spergel \& Pen 1997) and two quintessence models exhibiting a
variation in $w$ at small $z$ ($w(z)=-0.5+0.1\,z$ and
$w(z)=-0.5-0.05\,z$). In all cases we assumed a flat universe and
marginalised over $\Omega_M$ ($\Omega_M=0.3 \pm 0.1$). We also assumed that
redshifts of lensed objects are determined with an accuracy $\Delta z=0.3$
and that we have data out to redshift of $z_{\rm max}=5$.  This figure
shows that the strength of the constraints depends strongly on the fiducial
model. In fact, we found that fiducial models for which
$w_{\scriptscriptstyle 0}+ z(dw/dz)_{\scriptscriptstyle 0}\sim -1$ (for $z$
of order unity) give weaker constraints. This result is not surprising and
can be understood by simply using the fact that number of lensed objects
out to a redshift of $z$ is roughly proportional to the volume of the
universe $V(z)$. Since $dV/dp_i$ (where $p_i$ is either
$w_{\scriptscriptstyle 0}$ or $(dw/dz)_{\scriptscriptstyle 0}$) is
proportional to $(1+z)^{1+w_{\scriptscriptstyle 0}+z
(dw/dz)_{\scriptscriptstyle 0}}$, we see that the expected number of lensed
quasars varies slowly with $p_i$ if $w_{\scriptscriptstyle 0}+z
(dw/dz)_{\scriptscriptstyle 0}\sim -1$. In that case, our constraint region
will be relatively large.

\section{Summary \& Conclusions} 

In this {\it Letter}, we considered the possibility of constraining
quintessence models that have been suggested to explain the missing energy
density of the Universe.  We suggested gravitational lensing statistics,
which can be used as a probe of the equation of state of the missing
component, $w(z)$.  An exciting possibility to obtain an adequate sample of
lensed quasars and their redshifts comes from the Sloan Digital Sky
Survey. Writing $w(z)\approx w_{\scriptscriptstyle 0} + z
(dw/dz)_{\scriptscriptstyle 0}$, we studied the expected accuracy to which
equation of state today $w_{\scriptscriptstyle 0}$ and its rate of change
$(dw/dz)_{\scriptscriptstyle 0}$ can simultaneously be
constrained. Adopting some conservative assumptions about the quality of
the data from SDSS and assuming a flat universe with $\Omega_M=0.3 \pm
0.1$, we conclude that tight constraints on these two parameters can indeed
be obtained.  The strength of the constraints depends not only on the
quality of the lensing data from the SDSS, but also on the fiducial model
(true values of $w_{\scriptscriptstyle 0}$ and $(dw/dz)_{\scriptscriptstyle
0}$). In particular, fiducial models for which $w_{\scriptscriptstyle 0}+
z(dw/dz)_{\scriptscriptstyle 0}\sim -1$ (for $z$ of order unity) give
weaker constraints on $w_{\scriptscriptstyle 0}$ and
$(dw/dz)_{\scriptscriptstyle 0}$.

\acknowledgments

We would like to thank Scott Dodelson, Josh Frieman, Lloyd Knox, Cole
Miller, Jean Quashnock and Michael Turner for useful discussions. ARC
acknowledges support from McCormick Fellowship at the University of
Chicago. We also thank the anonymous referee for his/her prompt
refereeing of the paper.

\end{document}